\documentclass[10pt,conference]{IEEEtran} 

\usepackage{times}
\usepackage{multirow}
\usepackage{epsfig}
\usepackage{amsmath}
\usepackage{amssymb}
\usepackage{latexsym}
\usepackage{graphicx}
\usepackage{amsthm}
\usepackage{color}

\IEEEoverridecommandlockouts 

\begin{document}
\date{}

\title{\large{Speeding Multicast by Acknowledgment Reduction Technique }(SMART)}
\author{Arman Rezaee \thanks{This work is sponsored by the Department of Defense under Air Force Contract \# FA8721-05-C-0002. Opinions, interpretations, recommendations and conclusions are those of the authors and are not necessarily endorsed by the United States Government. Specifically, this work was supported by Information Systems of ASD(R\&E).} \vspace{-0.15cm}\\ \\Research Laboratory of Electronics \\ Massachusetts Institute of Technology\\Cambridge, MA.\\Email: armanr@mit.edu \and Linda Zeger \vspace{-0.15cm}\\ \\MIT Lincoln Laboratory\\Lexington, MA.\\Email: zeger@ll.mit.edu \and Muriel M\'{e}dard \vspace{-0.15cm} \\ \\Research Laboratory of Electronics\\  Massachusetts Institute of Technology\\Cambridge, MA.\\Email: medard@mit.edu}

\maketitle

\begin{abstract}
We present a novel feedback protocol for wireless broadcast networks that utilize linear network coding. We consider transmission of packets from one source to many receivers over a single-hop broadcast erasure channel. Our method utilizes a predictive model to request feedback only when the probability that all receivers have completed decoding is significant. In addition, our proposed NACK-based feedback mechanism enables all receivers to request, within a \textit{single} time slot, the number of retransmissions needed for successful decoding.
We present simulation results as well as analytical results that show the favorable scalability of our technique as the number of receivers, file size, and packet erasure probability increase. We also show the robustness of this scheme to uncertainty in the predictive model, including uncertainty in the number of receiving nodes and the packet erasure probability, as well as to losses of the feedback itself. Our scheme, SMART, is shown to perform nearly as well as an omniscient transmitter that requires no feedback. Furthermore, SMART, is shown to outperform current state of the art methods at any given erasure probability, file size, and numbers of receivers.

\end{abstract}

\section{Introduction} \label{sec:Intro}
Reliability is a challenging issue in wireless communications, particularly as the number of nodes becomes large, in which case conventional acknowledgment methods can result in unmanageable growth of feedback. We propose a new feedback mechanism for wireless broadcast networks and a predictive model that are built upon linear network coding. We dub our new approach Speeding Multicast by Acknowledgment Reduction Technique (SMART). The novelties of SMART are that it provides a predictive model for the time at which transmissions are likely to be able to be terminated and it also reduces the feedback from all users to one time slot per request. The primary relevant piece of information the transmitter would derive from the feedback is the number of degrees of freedom missing at the worst receiver. Combination of network coding and the predictive model allows the transmitter to use this information to substantially reduce the amount of feedback as well as unnecessary retransmissions.

Our proposed feedback mechanism has four main advantages over previous schemes: First, unnecessary initial polling by the transmitter is eliminated by use of the predictive model. Secondly, a significant reduction in the number of time slots allocated for feedback is achieved; this number currently scales with the number $n$ of receivers, but under the new method will become a scalar of order $1$. Thirdly, unnecessary retransmissions to the receivers can be greatly reduced. Fourthly, SMART is robust and is quite scalable, even to an uncertain number of receiving nodes. 

A prime example of an appropriate application of this method can be seen in large latency and delay challenged networks \cite{lucani:delay_challenged2009}, where feedback about received packets may be considerably delayed, reducing the feedback's usefulness and accuracy about the current state of the network. Other applications range from reliable bulk data transfer to streaming video to a large set of receivers. 

We study the performance gains of this feedback strategy, and compare it to the delay/throughput performance of an omniscient transmitter that requires no feedback. We also compare SMART to a wireless representation of a state of the art \textit{negative} feedback protocol, NACK-Oriented Reliable Multicast (NORM) \cite{Adamson:2009}. 

The rest of the paper is organized as follows: In Section \ref{sec:NetMod}, the network model and parameters are introduced. In Section \ref{sec:Mechanism}, we present the feedback mechanism. In Section \ref{sec:Disc_Model}, we evaluate the delay performance of the broadcast network under a discrete slotted model. In Section \ref{sec:Cont_Model}, we demonstrate the scalability of SMART to very large numbers of receivers. In Section \ref{sec:Robustness_SMART} we discuss the robustness of SMART to channel estimation errors, NACK erasures, and correlated losses. Section \ref{sec:Compare_w_NORM} compares the performance of SMART to a genie bound as well as to the NORM protocol. Finally, we provide a summary and concluding remarks in Section \ref{sec:Conclusion}.

\section{Network Model and Parameters}\label{sec:NetMod}
Consider a wireless broadcast scenario in which a node transmits a single file consisting of $k$ packets to $n$ independent users. In such systems a feedback mechanism is required to notify the transmitting node if all packets are received by the $n$ users or further transmissions are needed. The transmitting node could be a base station or a peer node within the network, but for simplicity we now consider that node a base station. A given channel between the base station and the $i^{th}$ user can be modeled as an erasure channel with parameter $p_{i}$, where $p_{i}$ is the packet erasure probability on that channel. Assume that channels are independent across time and across receivers and the base station is required to successfully complete the transmission of its packets to all $n$ users. We also assume that the base station uses network coding in the transmission of its packets, thus in the remainder of this paper we will use packets and degrees of freedom interchangeably. The analysis in Section \ref{sec:Predictive_Model} assumes that all channels are statistically identical and have the same packet erasure probability denoted by $p_{e}$, but it can be readily generalized to allow each channel to have its own distinct erasure probability.

\section{Feedback Mechanism}\label{sec:Mechanism}
The main idea that enables a single slot feedback is the use of CDMA codes. During the feedback slot, any receiver that has not correctly decoded the file will send a predetermined CDMA codeword to the base station, which indicates how many new degrees of freedom the base station needs to transmit for this user to recover all its missed packets. Two examples of CDMA codes are DS-CDMA and jitter.

With jitter, any of the $n$ users that have not correctly decoded all $k$ packets will send a short pulse to the base station, the timing of which indicates how many new degrees of freedom the user needs to decode the entire file. The feedback slot can be viewed as a concatenation of subslots whereby the presence of a pulse in a specific subslot will indicate that a corresponding predetermined number or percentage of dofs is needed. 

We propose the following scheme: the larger the number of degrees of freedom a receiving node will request, the earlier the subslot in which it will transmit within the single feedback slot. Thus, the base station will aim to find the first subslot in which a user transmits. If DS-CDMA were used, then the base station would first apply the matched filter corresponding to the highest percentage range of dofs requested. If a detection is found, the base station would be done processing the NACK slot, and would then transmit the highest number of dofs requested. If a detection is not found, the base station would next apply the matched filter corresponding to the second highest number of dofs, and the process is repeated. The ordering of CDMA codes would be chosen so that pairs of codes that represent similar percents of missing dofs would have higher correlations than pairs of codes that represent vastly differing percents of missing dofs. This ordering will increase the robustness of SMART to NACK erasures as well as to a noisy NACK channel. 
It should be noted that the single-slot mechanism is a physical layer enhancement, and a transport layer designer may not have control over it. However, the predictive model can be used to ensure that there will be feedback only from a minimal number of users, or the transmitter can transmit enough extra coded packets to ensure with high probability that there is feedback from at most one user.

If ordered CDMA codes or the associated receiver processing are not available to form this single slot feedback mechanism, having all NACKs transmitted in a single slot can still potentially be accomplished by other methods. For example, an energy detection mechanism at the base station can enable the base station to know whether or not all users have successfully received the file. The base station can then, according to the predictive model, select a time of feedback large enough so that the probability of any nodes needing more than one additional coded packet is small. As shown with our calculations presented in the next section, and confirmed by simulations, unless the erasure probability is large, the number of additional time slots needed to ensure this criteria is small.

\section{Predictive Model}\label{sec:Predictive_Model}
\begin{figure} [ht]
\centering
\vspace{-0.2cm}
\includegraphics[scale=0.43,clip=true]{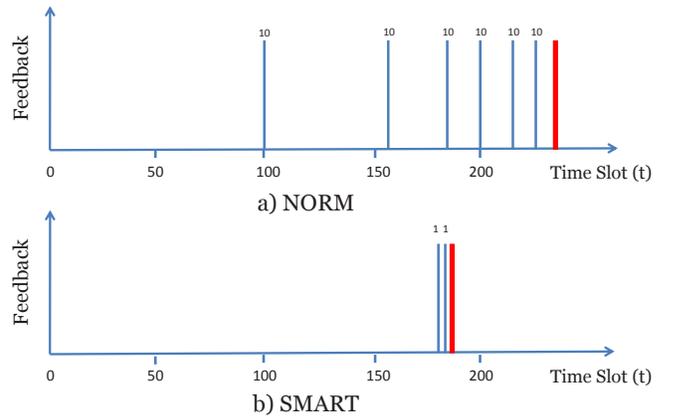}
\caption{Feedback times for NORM vs. SMART for $n = 10000$, $k = 100$, and $p_{e} = 0.3$. The number above each blue bar indicates the number of slots devoted to NACKs at that cycle and red bars denote the end of transmissions.}
\label{fig:compare_predictions}
\end{figure}

In this section, we demonstrate the prediction capability of our method and we show that the receivers should be polled if and only if there is a reasonable probability that they have completed their download. Fig. \ref{fig:compare_predictions} captures the difference between traditional protocols and our scheme by showing sample feedback times of both mechanisms. With currently available protocols, including NORM enhanced by feedback suppression, NACKs occupy a proportion of the slots throughout the transmission. In contrast, our SMART scheme allows for strategic placement of the NACKs at only a few isolated slots near the download completion time. SMART considers the inherently lossy nature of the channel and incorporates the predicted loss in scheduling of the feedback. Furthermore, each feedback cycle of SMART utilizes only a single slot, whereas NORM utilizes multiple slots, for example 10, as shown in Fig. \ref{fig:compare_predictions}.

\subsection{Performance evaluation in a Discrete Model}\label{sec:Disc_Model}
In this section, we analyze the number of time slots needed to reliably transmit a file of $k$ packets to $n$ receivers. For example, the $k$ coded packets could represent a file or image. We consider a slotted broadcast channel where each transmitted packet is received independently with probability $1 - p_e$ at any of the $n$ receivers, where $p_e$ is the packet erasure probability. This model is equivalent to $n$ independent Bernoulli processes, each with parameter $1-p_e$, where we are interested in the shortest time until all processes have had $k$ successful arrivals. The case of correlated users will be discussed in \ref{sec:Robustness_SMART}.

The transmission is completed when each of the $n$ receivers has successfully received $k$ or more coded packets. Let us denote the number of degrees of freedom (dof) missing at node $i$ after $t$ time slots by $M_{i}^{t} \in [0,k]$. We define another random variable $M_{t} = \max \{M_{1}^{t}, M_{2}^{t},...M_{n}^{t} \}$ to denote the number of dofs missing at the node that has experienced the highest number of erasures during $t$ transmissions. Ideally, the transmitter should receive feedback enabling it to stop at $\{\min(t) | M_{t} = 0 \}$. The probability that receiver $i$ has received $k$ or more coded packets in $t$ time slots is:
\begin{eqnarray}
Pr\{M_{i}^{t} = 0\} & = & 1 - \sum_{j=0}^{k-1} {t \choose j} p_{e}^{t-j} (1-p_{e})^{j}
\end{eqnarray}

Similarly, let us denote the probability that all $n$ receivers have completed the download after $t$ time slots by $\beta(t)$:
\begin{eqnarray}
\beta(t) & = & Pr\{M_{t} = 0\}  =  \left( Pr \left \{M_{i}^{t} = 0 \right \}\right)^n  \nonumber \\ 
& = & \left ( 1 - \sum_{j=0}^{k-1} {t \choose j}  p_{e}^{t-j} (1-p_{e})^{j} \right)^n \label{eq:Completion}
\end{eqnarray}

\begin{figure} [ht]
\centering
\vspace{-0.2cm}
\includegraphics[scale=0.48,clip=true]{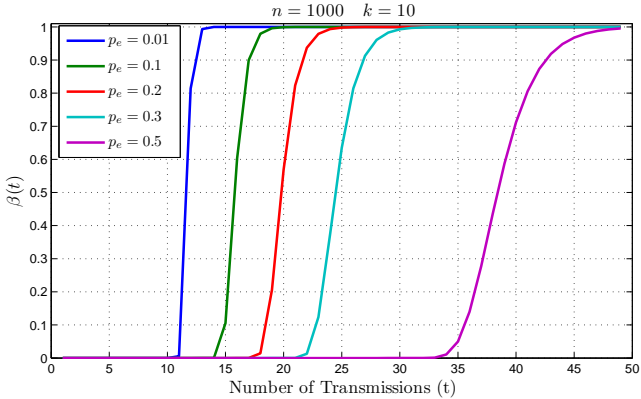}
\caption{Completion probability as a function of $t$ for different values of $p_e$.}
\label{fig:diffP}
\end{figure}

Note that $\beta(t)$ is the probability that transmissions can cease after $t$ time slots. In the following figures, we show how $\beta(t)$ changes as a function of $p_e, k,$ and $n$. Fig. \ref{fig:diffP} depicts $\beta(t)$ for a range of erasure probabilities. Notice that the time at which transmissions can cease is very sensitive to packet erasure probability. As shown, for a network of $n = 1000$ nodes and $k = 10$ packets, the probability $\beta(t) = 0.7$ is achieved after 21 time slots when $p_{e} = 0.2$. This number increases to 40 time slots when $p_{e} = 0.5$. An important feature of this graph is the shape of the $\beta(t)$ function; the probabilities rise more sharply for smaller erasure probabilities than for larger ones. Simulations of SMART show that for small $p_{e}$ the protocol will have minimal number of feedback cycles (an average of slightly more than 1 cycle), while for large $p_e$ at most a few more cycles are needed. 

\begin{figure} [ht]
\centering
\vspace{-0.2cm}
\includegraphics[scale=0.48,clip=true]{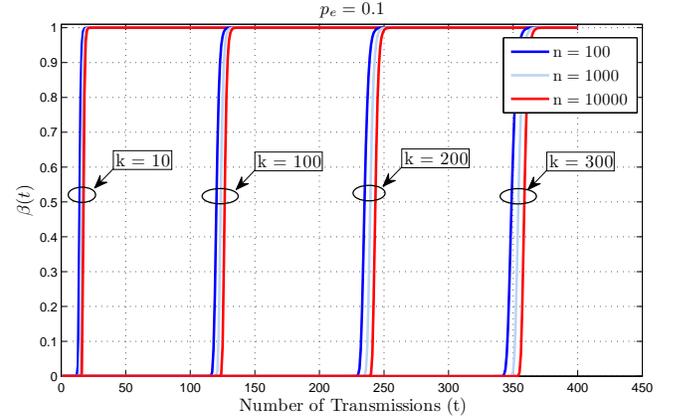}
\caption{Completion probability as a function of $t$ for varying $k$ and $n$.}
\label{fig:diff_k_and_n}
\end{figure}

Fig. \ref{fig:diff_k_and_n} shows the number of transmissions required to achieve a specified reliability, as we scale the size of the file and the number of receiving nodes. Notice that the number of transmissions is strongly dependent on the file size $k$ but is not very sensitive to the number of receivers $n$. As shown, doubling the number of packets in the file will roughly double the number of transmissions needed for any given reliability. Ovals within the figure are used to show the proximity of the curves that correspond to an increase in $n$ (from $100$ to $10000$) for a fixed $k$. As we show analytically in Section \ref{sec:Cont_Model}, the number of coded packet transmissions required for a given reliability is not very sensitive to an increase in $n$, so as seen in Fig. \ref{fig:diff_k_and_n}, large changes in $n$ requires small changes to the number of packet transmissions. The figure shows the robustness of this transmission scheme to uncertainty in $n$.
It is important to note that in both figures, the CDFs have very sharp increases when erasure probability is not too high: a significant increase in reliability is achieved by very few extra transmissions. Therefore, SMART avoids extraneous transmissions and feedbacks by only requesting feedback when $\beta(t)$ is sufficiently large; the base station will then be able to cease transmissions $\beta(t)$ proportion of the time. 

Once we have scheduled the initial feedback at time $t$, we are interested in the number of nodes that will then be requesting feedback in the single slot. We thus calculate the expected value of the minimum of $n$ random variables. Let us denote the number of nodes that have not completed the download at time $t$ by a random variable $N_{1}$ and also use $\overline{N_{1}}$ as its expected value. The probability mass function (pmf) of $N_{1}$ is:
\begin{eqnarray}
\small Pr \left\{  N_{1} = i \right\} 
& = & \small {n \choose i}  \left( Pr \small \left\{ \begin{array}{l} 
\mbox{1 node completed} \\ 
\mbox{the download by t} 
\end{array} \right\} \right)^{n-i} \nonumber \\
& \cdot & \left( \small 1 - Pr \small \left\{ \begin{array}{l} 
\mbox{1 node completed} \\ 
\mbox{the download by t} 
\end{array} \right\} \right)^{i} \label{eq:pmf_of_N1}
\end{eqnarray}

\begin{figure} [ht]
\centering
\vspace{-0.2cm}
\includegraphics[scale=0.48,clip=true]{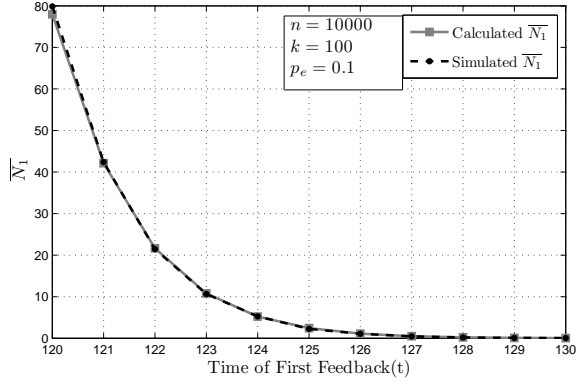}
\caption{Expected number of nodes that have not completed the download.}
\label{fig:sim_vs_calc_N1}
\end{figure}

Recall that $N_{1}$ is non-negative, thus:
\begin{eqnarray}
\overline{N_{1}} = \int^{\infty}_{0} \left( 1 - F(x) \right) dx \label{eq:average_N1}
\end{eqnarray}
where $F$ is the cumulative distribution function of $N_{1}$. Fig. \ref{fig:sim_vs_calc_N1} depicts $\overline{N_{1}}$ for different initial feedback times for a network of $n = 10,000$ nodes with $k = 100$ packets and $p_{e} = 0.1$. Notice that the analytical results of (\ref{eq:average_N1}) and the simulated values are virtually identical and $\overline{N_{1}}$ decreases rapidly with the initial feedback time. Smaller values of $\overline{N_{1}}$ are particularly helpful in reducing feedback traffic if the single-slot mechanism is difficult to fully implement.

After feedback, the base station will know the number $M_{t}$ of packets missing at the worst receiver. We allow the transmitter to transmit $\widehat{M}_{t} = \frac{M_{t}}{1-p_{e}}$ coded packets after the feedback. We can calculate the probability $\widehat{\beta}$ that everyone has completed the download $\widehat{M}_{t}$ time slots after the first feedback, given that $\overline{N_{1}}$ nodes have not completed the download by the time $t$ of the initial feedback.
\begin{eqnarray}
\widehat{\beta} (\widehat{M}_{t}) & = &
Pr \left\{ \begin{array}{l} 
\mbox{ everyone completed} \\ 
\mbox{ the download by $t + \widehat{M}_{t}$} 
\end{array} \right\} \nonumber \\ 
& = & Pr \left\{ \begin{array}{l} 
\mbox{$\overline{N_{1}}$ nodes completed} \\ 
\mbox{the download in $\widehat{M}_{t}$ slots} 
\end{array} \right\}  \label{eq:upper_bound} 
\end{eqnarray}
Following $\widehat{M}_{t}$ transmissions by the base station, another single slot of feedback is allocated, and if any NACKs are received in it, the process is repeated. Simulations have shown that the entire file download is usually accomplished within at most a few cycles of the first feedback.

\subsection{Performance evaluation in a Continuous Model}\label{sec:Cont_Model}
In this section, we will derive the scaling laws for the performance of the system when transmissions are modeled as continuous. We model the arrivals at each receiver as a Poisson process and analyze the behavior of completion time as the number of receivers $n$ grow.

Each of the $n$ users needs to receive $k$ or more coded packets from a single transmitting node. In time $t$ packet lengths, each of the $n$ nodes \textit{independently} receives a number of packets that is Poisson distributed, on the time scale of integral numbers of packet lengths, with parameter $\lambda t$, where $\lambda = 1 - p_e$, and $p_e$ is the packet erasure probability. The probability that user $i$ receives $k$ or more coded packets within time $t$ is thus:
\begin{eqnarray}
Pr\{M_{i}^{t} = 0\} & = & 1 - \sum_{j = 0}^{k-1} \frac{(\lambda t)^j \exp(-\lambda t)}{j!} \label{one-user}
\end{eqnarray}

Hence the probability that all $n$ users receive at least $k$ coded packets in time $t$ or earlier is (\ref{one-user}) raised to the power of $n$. As in Section \ref{sec:Disc_Model} we define $\beta(t)$ to be the probability that \textit{all} of the $n$ users received $k$ or more coded packets within time $t$. This probability $\beta(t)$, which is also the probability that the transmitter can stop sending coded packets, is:
\begin{eqnarray}
\beta(t) & = & \left(1 - \sum_{j = 0}^{k-1} \frac{(\lambda t)^{j} \exp(-\lambda t)}{j!} \right)^{n} \label{eq:cont_model}
\end{eqnarray}

We select the first feedback time so that there is a significant probability that every receiver has completed the download and there is no need for retransmissions. In other words, $t^{*}$ is a time whose corresponding $\beta(t^*)$ has reached a certain reliability threshold. Let us use $\beta^{*}$ to denote this threshold. Thus: 
\begin{eqnarray}
t^{*} & = & t^{*}\big(\beta^{*},n\big) = \inf \left \{\hspace{0.05cm} t \hspace{0.2cm} | \hspace{0.2cm} \beta(t) \geq \beta^{*} \right \}
\end{eqnarray}

Rearranging terms in (\ref{eq:cont_model}) and substituting $\beta^{*}$ and $t^{*}$ for $\beta(t)$ and $t$ yields:
\begin{eqnarray}
\lambda t^{*} & = & \ln \left (\sum_{j = 0}^{k-1} \frac{(\lambda t^{*})^j}{j!}\right) - \ln \left( 1 - {\beta^{*}}^\frac{1}{n} \right) \label{exact-expression} \\
& = & \lambda t^{*} + \ln \left(\frac{\Gamma(k,\lambda t^{*})}{\Gamma(k)}\right) - \ln \left( 1 - {\beta^{*}}^\frac{1}{n} \right)
\end{eqnarray}

we then have:
\begin{eqnarray} \label{gamma_eqn}
\frac{\Gamma(k,\lambda t^{*})}{\Gamma(k)} & = &  \left( 1 - {\beta^{*}}^\frac{1}{n} \right)  \label{eq:exact_sensitivity_parameters}
\end{eqnarray}
where the Gamma functions are defined as:
\begin{eqnarray}
\Gamma(a,b) = \int_{b}^{\infty} t^{a-1} e^{-t} dt \nonumber \\ 
\Gamma(a) = \int_{0}^{\infty} t^{a-1} e^{-t} dt \nonumber 
\end{eqnarray}

\begin{figure} [ht]
\centering
\includegraphics[scale=0.63,clip=true]{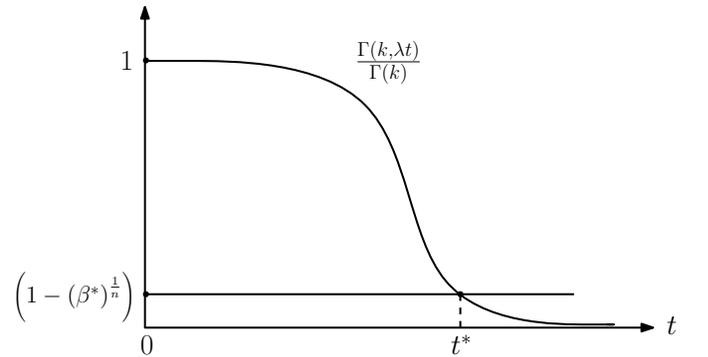}
\caption{Calculating $t^{*}$ from the $\frac{\Gamma(k,\lambda t)}{\Gamma(k)}$ function.}
\label{fig:Gamma_function_for_cont_model}
\end{figure}
Fig. \ref{fig:Gamma_function_for_cont_model} illustrates equation (\ref{gamma_eqn}). Notice that $\frac{\Gamma(k,\lambda t)}{\Gamma(k)}$ is strictly decreasing in $t$ and is thus invertible. As a result, given a set of parameters $(n,k,\beta^{*})$, a unique $t^{*}$ can be determined that is the amount of time that it takes for all $n$ users to receive the $k$ packet file, with probability $\beta^{*}$. The right hand side of (\ref{gamma_eqn}) corresponds to the horizontal line in Fig. \ref{fig:Gamma_function_for_cont_model}, and is the probability that any given user has not received the file by time $t^{*}$. For large $n$ and even a modest $\beta^{*}$, this probability, and hence the resulting horizontal line, would be quite low, resulting in the selection of a $t^{*}$ such as that shown in the figure. Alternatively, if the function $\frac{\Gamma(k,\lambda t)}{\Gamma(k)}$ is considered at time $t$, rather than at time $t^*$, then raising $1 - \frac{\Gamma(k,\lambda  t)}{\Gamma(k)}$ to the power of $n$ yields a continuous model version of the the probability function $\beta(t)$ plotted in Fig. \ref{fig:diff_k_and_n}. Taking the $n^{th}$ power of $1 - \frac{\Gamma(k,\lambda  t)}{\Gamma(k)}$ for large $n$ renders $\beta(t)$ close to 1 only if $1 - \frac{\Gamma(k,\lambda  t)}{\Gamma(k)}$ is very close to 1, thereby yielding the sharp transition in time seen in Fig. \ref{fig:diff_k_and_n}.

We are interested in sensitivity of $t^{*}$ to $n$ for a given value of $\beta^{*}$. A better understanding of this sensitivity can be achieved by looking at the reverse problem. Let us see how many nodes $n$ we can accommodate after $t$ transmissions for a given value of $\beta^{*}$. Rearranging terms in (\ref{eq:exact_sensitivity_parameters}) and solving for $n$ yields: 
\begin{eqnarray}
n & = & \frac{\ln \big(\beta^{*}\big) }{\ln \Big( 1 - \frac{\Gamma(k,\lambda t)}{\Gamma(k)} \Big)} \label{eq:exact_sensitivity_for_figures}
\end{eqnarray}

\begin{figure} [ht]
\centering
\includegraphics[scale=0.62,clip=true]{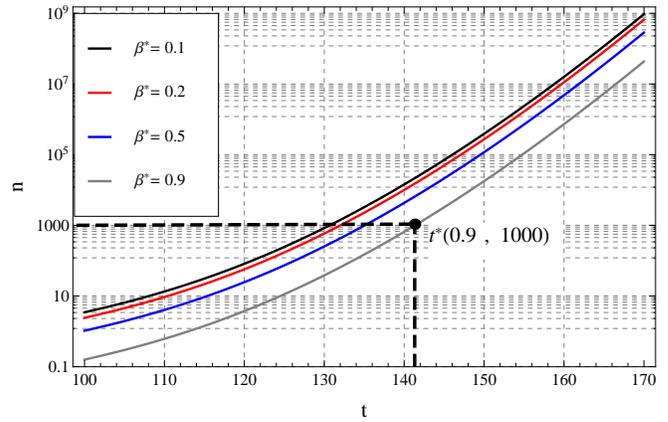}
\vspace{-0.8cm}
\caption{The number of nodes $n$ that can be accommodated for a given transmission time $t$. The figure was computed according to (\ref{eq:exact_sensitivity_for_figures}) with $k=100$ packets and $p_{e} = 0.1$. An example of how the time $t^{*}\big(\beta^{*},n\big)$ can be obtained from these curves is illustrated by the dashed lines for the case of $\beta^{*}$ = .9 and $n = 1000$.}
\label{fig:accomodated_nodes_for_a_given_t_k_100}
\end{figure}

Figure \ref{fig:accomodated_nodes_for_a_given_t_k_100} provides the number of users that can be accommodated by time $t$, for a range of $\beta^{*}$. The figure was computed according to (\ref{eq:exact_sensitivity_for_figures}) for a file size of $k = 100$ packets and packet erasure probability of $p_{e} = 0.1$. Fig. \ref{fig:accomodated_nodes_for_a_given_t_k_100} can be used to determine the $t^{*}$ that will ensure a given reliability $\beta^{*}$ for a given $k$ and $n$. The dashed black lines in the figure illustrate how to determine this time for the example case of $\beta^{*}$ = .9 and $n = 1000$.

The number of nodes $n$ that can be accommodated increases rapidly, as emphasized by the logarithmic scale of the vertical axis and the linear scale of the horizontal axis. In fact a much larger group of users can be accommodated with a relatively short extra transmission time. For example, when $\beta^{*} = 0.1$, an increase of approximately $20$ in $t$ (from $110$ to $130$) can accommodate $100$ times as many users (from 10 to 1000 users). Because of the convexity exhibited in the figure, ever larger groups can be accommodated with the same number of extra transmissions. 

It should also be noted that $n$ is not very sensitive to $\beta^{*}$, and the sensitivity decreases as $t$ increases. For example, the figure shows that in order to accommodate $n = 10$ users, with reliabilities $\beta^{*} = 0.1$ and $\beta^{*} = 0.9$, we need $t = 108$ and $t = 125$ respectively (a 15.7\% increase in $t$ to reach the higher $\beta^{*}$). Accommodating $n = 1000$ users for the same values of $\beta^{*}$ will require $t = 130$ and $t = 142$ respectively (a 9.2\% increase in $t$). 

Similar numerical results hold for larger file sizes and can be verified by plotting (\ref{eq:exact_sensitivity_for_figures}) for larger values of $k$. As $k$ increases, the \textit{per packet} time required to reliably transmit a file to a fixed number of receivers decreases. This favorable gain comes from the ability to code across larger files, and shows the robustness of SMART to increases in the file size.


\subsection{Robustness of SMART}\label{sec:Robustness_SMART}
Robustness of SMART to channel estimation errors is mainly the result of its single slot characteristic. If physical considerations do not allow for an accurate estimation of the channel, an appropriately conservative approach is to underestimate $p_{e}$ so that the predictive model will schedule the initial feedback at an earlier time slot. Since the feedback penalty is only $1$ time slot, the earlier feedback will avoid significant loss of throughput and we can adjust the previous estimation of $p_{e}$ based on the feedback. Simulation results show that for a network of $n = 1000$ receivers and $k = 100$ packets if a channel with $p_{e} = 0.2$ was estimated to have $p_{e} = 0.1$ the total download time will be increased from $151$ to $152$ time slots. 

SMART is also robust against correlated losses among users. Correlation of erasures among users can be thought of as reducing $n$, the number of independent users, and thus will have a similar effect to decreasing $n$. We showed in \ref{sec:Cont_Model} the total download time is not very sensitive to $n$, and thus correlation is not expected to affect the results substantially in most cases.

Robustness of SMART to NACK erasures is also superior to other protocols. Unlike NACK suppression schemes that allow only a few nodes to send their feedback, SMART allows every eligible node to participate in the feedback and if a NACK is erased, the base station will be able to use the feedback from other nodes. As an additional robustness feature, if the base station does not receive any NACKs during a feedback cycle, another feedback slot will be scheduled immediately to confirm that transmissions can end. This increases robustness to NACK erasures with minimal cost to total download time.

\section{Comparison to Other Multicasting Protocols} \label{sec:Compare_w_NORM}
We performed simulations of SMART over a range of $k$, $n$, $p_e$, and $\beta^{*}$. The simulations showed that while the value of $\overline{N_{1}}$, as plotted in Fig. 4, as well as the number of outstanding packets needed, varies with the time of the first feedback, the total completion time was generally not sensitive to the precise value of $t^{*}$ used. 

\begin{figure} [ht]
\centering
\includegraphics[scale=0.48,clip=true]{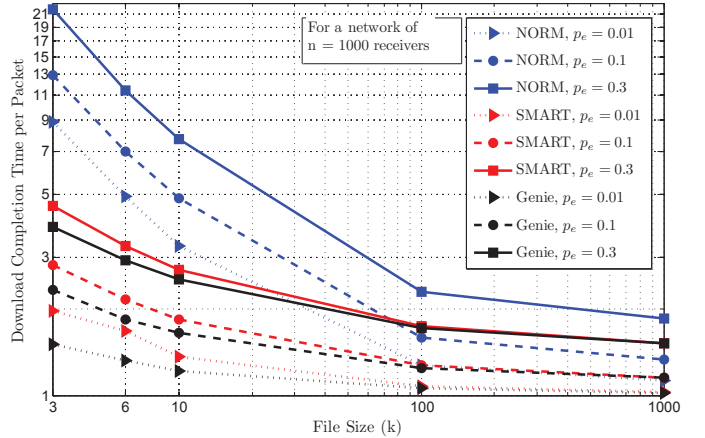}
\caption{Simulation results depicting the performance of SMART, the theoretical bound obtained from a genie based protocol, and a wireless representation of NORM.}
\label{fig:SMART_vs_NORM}
\end{figure}

The red curves of Fig. \ref{fig:SMART_vs_NORM} plot on a log-log scale the download completion time \textit{per packet} of SMART vs. file size $k$, for a network of $n = 1000$ receivers. Recall that with SMART the total download time is not very sensitive to $n$ and the SMART curves in Fig. \ref{fig:SMART_vs_NORM} will thus change only slightly as the network size increases.

The theoretical genie bound, in which the base station always knows how many coded packets each receiver is missing without any transmitted feedback, is shown in black in Fig. \ref{fig:SMART_vs_NORM}. It is seen that SMART performs almost as well as such an omniscient base station that requires no feedback, particularly at larger file sizes. This behavior occurs because the number of slots allocated for feedback in SMART will stay approximately constant regardless of the file size or the erasure probability.

The blue curves represent the performance of a wireless representation of NORM. NACK-based protocols such as NORM \cite{Adamson:2009} have been proposed to provide end-to-end reliable transport of bulk data while avoiding the feedback implosion associated with reliable multicast. In order to reduce the amount of feedback, NORM, like SMART, utilizes negative acknowledgments (NACKs), rather than the positive acknowledgments (ACKs) used by earlier protocols. NORM also uses end-to-end coding, which is equivalent to network coding for the single hop example illustrated here. End-to-end coding incurs a longer time for each feedback cycle, which we did not include in our representation of the NORM model. Furthermore, our single slot feedback mechanism relies on the base station that receives the wireless nodes' feedback in a single slot to process this feedback, and adjust or terminate its transmissions of coded packets accordingly.

While we have attempted to select representative modes and settings of NORM and to optimistically model its performance in a wireless setting, it is possible that other choices of parameters could provide better performance. A central feature of NORM is its \emph{NACK-suppression} scheme \cite{Adamson:2002}. In NORM's default setting, FEC is sent \textit{only} in response to NACKs and according to \cite{Adamson:2008}, the base station allocates between $5$ to $7$ round trip times to NACK aggregation before restarting the transmission, which is equivalent to $10$-$14$ time slots.

We have assumed that NORM spends $10$ time slots for NACK aggregation during each feedback cycle and experiences no NACK collisions at the base station. We also model the Reed-Solomon (RS) coding option of NORM \cite{Luby:2002}; if $k < 250$ packets, the entire file is considered as a single RS block, in which case exactly $k$ successful packet receptions are required for decoding. For larger file sizes, we approximate NORM as using a series of $250$-packet RS blocks, and the transmitter will move on to the second RS block if and only if the first block is decoded at all receivers. A block size smaller than $256$ packets was recommended by \cite{Luby:2002} to avoid high decoding complexity. 

As shown, SMART outperforms NORM at every erasure probability and for any file size. In particular, note that NORM's performance is detrimentally affected when the file size is small, which occurs because the penalty associated with the NACK aggregation wait dominates over the data transmission time. As shown in the figure, SMART's per packet completion time is very close to 1 for large files. In contrast, for files of greater than 250 packets NORM is seen to have a larger constant download time per packet. For large files, network coding overhead of SMART resulting from encoding of the coefficients can be prevented if we initialize the random number generators at the transmitter and receivers with the same seed \cite{JayKumar:NetworkCoding_MeetsTCP}.

\section{Conclusion}\label{sec:Conclusion}
We proposed a predictive model to determine suitable feedback times that will reduce the feedback traffic as well as transmission of extraneous coded packets in a broadcast erasure channel. We also introduced a novel \textit{single slot} feedback mechanism, that enables any number of receivers to simultaneously transmit their feedback. We showed the scalability of the combined predictive model and feedback mechanism with increasing file size, as well as with a large number of receivers. We showed the robustness of SMART, and we demonstrated that SMART's performance is close to that of an omniscient transmitter with no feedback. While homogeneous channels among the different receivers were discussed in this paper, ongoing work is considering channels with different erasure probabilities.

\bibliographystyle{IEEEtran}
\bibliography{IEEEabrv,references}

\end{document}